\documentclass[conference,a4paper]{IEEEtran}
\usepackage{cite}
\usepackage{bm}
\usepackage{float}
\usepackage{array}
\usepackage{multirow}
\usepackage[cmex10]{amsmath}
\usepackage{graphicx}
\usepackage{url}
\usepackage{color}
\usepackage{amssymb}
\usepackage{amsfonts}
\usepackage{amsmath}
\usepackage{extarrows}
\usepackage{graphicx}
\usepackage{amsfonts}
\usepackage{caption}
\usepackage{algorithm}
\usepackage{algpseudocode}
\usepackage{indentfirst}
\usepackage{caption}
\usepackage{esint} 
\usepackage{graphicx}
\usepackage{graphics}
\usepackage{pgfplots}
\usepackage{tikz}
\usepackage{epsfig}
\usepackage{soul}
\usepackage{balance}
\usepackage{caption}
\usepackage{gensymb}
\def\bA{{\mathbf{A}}}  \def\bC{{\mathbf{C}}}  
\def\bF{{\mathbf{F}}}  \def\bH{{\mathbf{H}}} \def\bI{{\mathbf{I}}} 
   \def\bN{{\mathbf{N}}} 
  \def\bR{{\mathbf{R}}}  
   \def\bX{{\mathbf{X}}} \def\bY{{\mathbf{Y}}}

\def\ba{{\mathbf{a}}}    
\def\bf{{\mathbf{f}}} \def\bg{{\mathbf{g}}} \def\bh{{\mathbf{h}}}  
   \def\bn{{\mathbf{n}}} 
   \def\bs{{\mathbf{s}}} 
   \def\bx{{\mathbf{x}}} \def\by{{\mathbf{y}}}
\def\bz{{\mathbf{z}}}

\IEEEoverridecommandlockouts
\begin{document}
\title{RIS-Augmented Millimeter-Wave MIMO Systems for Passive Drone Detection} 

\author{\IEEEauthorblockN{Jiguang~He$^1$,
Aymen~Fakhreddine$^{1,2}$, and George C. Alexandropoulos$^{3}$}
\IEEEauthorblockA{$^1$Technology Innovation Institute, 9639 Masdar City, Abu Dhabi, United Arab Emirates}
\IEEEauthorblockA{$^2$Institute of Networked and Embedded Systems, University of Klagenfurt, Klagenfurt, Austria}
\IEEEauthorblockA{$^3$Department of Informatics and Telecommunications, National and Kapodistrian University of Athens\\
Panepistimiopolis Ilissia, 15784 Athens, Greece}
\IEEEauthorblockA{E-mails: \{jiguang.he, aymen.fakhreddine\}@tii.ae, alexandg@di.uoa.gr}
}


 \maketitle
\begin{abstract}
In the past decade, the number of amateur drones is increasing, and this trend is expected to continue in the future. The security issues brought by abuse and misconduct of drones become more and more severe and may incur a negative impact to the society. In this paper, we leverage existing cellular multiple-input multiple-output (MIMO) base station (BS) infrastructure, operating at millimeter wave (mmWave) frequency bands, for drone detection in a device-free manner with the aid of one reconfigurable intelligent surface (RIS), deployed in the proximity of the BS. We theoretically examine the feasibility of drone detection with the aid of the generalized likelihood ratio test (GLRT) and validate via simulations that, the optimized deployment of an RIS can bring added benefits compared to RIS-free systems. In addition, the effect of RIS training beams, training overhead, and radar cross section, is investigated in order to offer theoretical design guidance for the proposed cellular RIS-based passive drone detection system.  
\end{abstract}

\section{Introduction}
In the literature, drone detection, recognition, localization, and tracking has been intensively studied, e.g., via computer vision (CV), acoustic arrays, radio frequency (RF) fingerprinting, and millimeter wave (mmWave) radar systems~\cite{Ismail2018}. For instance, by equipping high-definition cameras on the roof of the buildings, we can detect the absence/presence of a flying drone, in the sky. However, such a CV-based drone detection system suffers from poor weather and air conditions, e.g., fog, dust, and rain. Also, it can hardly distinguish an object with the same/similar shape as drones. The acoustic arrays suffer from background noise, and the detection range is rather limited due to the severe propagation path loss, usually less than 1000 meters. The RF fingerprint only works when the RF signals are emitted from either the drones for data streaming or the remote controller for control signaling. Data fusion can be exploited by fusing data from different sensors readings, but it inherently increases the deployment cost, the computational complexity, and maintenance expenses. 

Recently, there are ongoing activities under the IEEE 802.11bf on wireless sensing, which include absence/presence detection of human being, gesture recognition, object tracking, based on the received WiFi signals~\cite{restuccia2021ieee}. However, the activity is mainly constrained to the indoor scenario. The sensing capability of the cellular system, especially for drone detection, in an outdoor environment has not yet been investigated for the case where the transmitter and receiver are spatially distributed. In the literature, some works carried out the drone detection with reconfigurable intelligent surfaces (RISs). However, the transmitter and receiver are co-located, working as a monostatic radar and thus requiring a full-duplex operation~\cite{buzzi2021radar,Buzzi2022}. The introduction of cost-effective RISs \cite{huang2019reconfigurable,Marco2019} can boost the performance of these newly introduced sensing functionalities due to the following reasons: i) RIS can be used as a reference node (a.k.a. anchor)~\cite{He2020,RISaccessfree}; ii) RIS can create a virtual link/route between the base station (BS) and user equipment (UE), especially when direct BS-UE link is blocked~\cite{Elzanaty2021}; iii) RIS have multiple operation modes, which can be customized for not only communications, but also for sensing~\cite{Huang2020,RISisac}. 

In the paper, we extend the study from indoor scenario to outdoor scenario, and introduce RIS for performance enhancement of drone detection. We study the passive drone detection with the assistance of current cellular multiple-input multiple-output (MIMO) BS along with a passive RIS via the generalized likelihood ratio test (GLRT). The prior location information on the network nodes is leveraged for the training beam design at the BS. Thanks to the line-of-sight (LoS) availability between BS and RIS, a directional beam in the form of array response vector, i.e., maximum ratio transmission (MRT), is used for each time slot during the whole sounding procedure as to provide high beamforming gain for the path via the RIS. We examine the effect of different system parameters, e.g., the type of RIS training beams, the training overhead, and radar cross section (RCS), on the probability of detection. The numerical results verify that the proposed RIS-augmented mmWave MIMO systems outperform their RIS-free counterparts regarding drone detection performance.

\textit{Notations}: A bold lowercase letter $\ba$ denotes a vector, and a bold capital letter $\bA$ denotes a matrix. $(\cdot)^\mathsf{T}$ and $(\cdot)^\mathsf{H}$ denote the matrix or vector transpose and Hermitian transpose, respectively. $\bA^\dag$ denotes Moore–Penrose inverse of $\bA$. $\mathrm{diag}(\ba)$ denotes a square diagonal matrix with the entries of $\ba$ on its diagonal, $\mathrm{vec}(\bA)$ returns a vector by stacking the $\bA$'s columns on top of one another, $\mathrm{Tr}(\bA)$ denotes the trace of $\bA$, 
$\bI_{M}$ denotes the $M\times M$ identity matrix, $j = \sqrt{-1}$, and 
$\|\cdot\|_2$ denotes the Euclidean norm of a vector.

\section{System Model}
\begin{figure}[t]
	\centering
\includegraphics[width=0.85\linewidth]{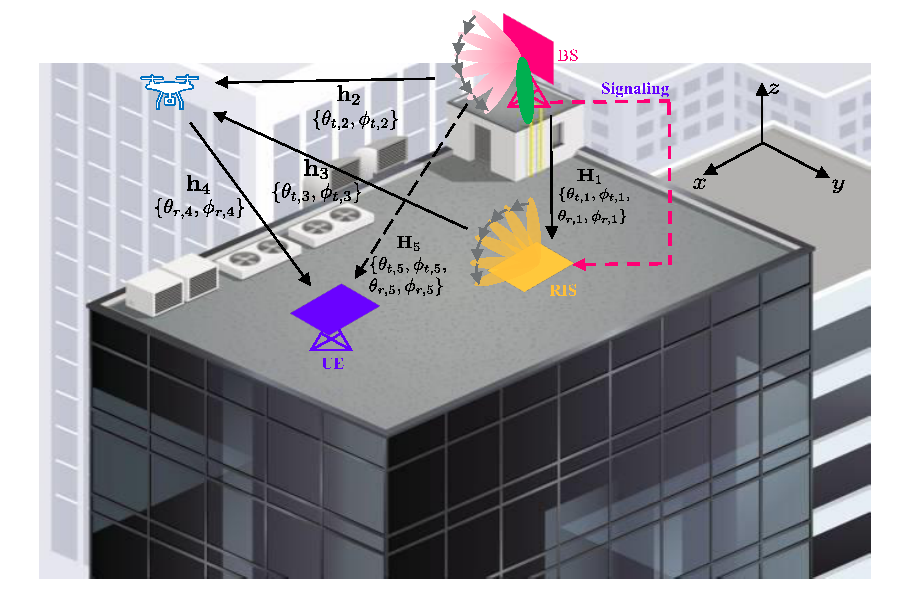}
	\caption{Drone detection system with one MIMO BS, one RIS, and one UE, all located on the roof of a building.}
		\label{System_Model}
 		\vspace{-0.5cm}
\end{figure}

The proposed drone detection system, shown in Fig.~\ref{System_Model}, is placed on the roof of a building, which includes one cellular MIMO BS, one passive RIS (i.e., 
no RF chains and baseband processing units are needed at the RIS), and one user equipment (UE), all equipped with multiple antennas (elements). 
A signaling link is established from the BS to the RIS, either via wireless or wired connection, for the purpose of coordination of RIS training beam design during the pilot transmission phase. The BS, the RIS, and the UE have an uniform planar/rectangular array (UPA/URA) structure. To be specific, the BS antenna array is parallel to the $y\text{-}z$ plane, while the RIS and UE arrays are parallel to the $x\text{-}y$ plane. Upon the deployment of the RIS, we assume that its exact position/location is known to the other network nodes, i.e., BS and UE. By leveraging this prior information, the purpose of such a system is to detect a flying drone with relatively low RCS, low velocity, and low altitude. This drone detection system is passive, also known as device-free, since the drones are not required to directly communicate and exchange information with the BS/UE.  
 
\subsection{Channel Model}
In general, the RIS can be placed in the LoS path of the BS, and the distance between them can be potentially chosen to be small in order to have a relatively small path loss. The channel between the BS and the RIS $\bH_1\in\mathbb{C}^{M_\text{R}\times M_\text{B}}$ with $M_\text{R}$ and $M_\text{B}$ being the numbers of RIS elements and BS antennas can be modeled by following the commonly-used Saleh-Valenzuela channel model, as~\cite{Heath2016} 
 \begin{align}\label{H_1}
\bH_1 =& \frac{e^{-j 2\pi  d_1/\lambda}}{\sqrt{\rho_1}} \boldsymbol{\alpha}_x(\theta_{r,1},\phi_{r,1}) \otimes \boldsymbol{\alpha}_y(\theta_{r,1},\phi_{r,1})\nonumber\\ &\Big(\boldsymbol{\alpha}_y(\theta_{t,1},\phi_{t,1}) \otimes \boldsymbol{\alpha}_z(\phi_{t,1})\Big)^\mathsf{H},
\end{align}
 where $\rho_1$ and $d_1$ are the path loss and distance between the cellular BS and the RIS,\footnote{Even though the two terms, i.e., $\rho_1$ and $d_1$, are closely coupled with one being a function of the other, we would like to define them separately for the sake of simplicity.} $j = \sqrt{-1}$, $\lambda$ is the wavelength of the carrier frequency, $\theta_{r,1}$ ($=\theta_{t,1}$) and $\phi_{r,1}$ ($=\phi_{t,1}$) are the azimuth and elevation angles of arrival (departure) associated with $\bH_1$, respectively. Here, $t$ and $r$ in the subscripts indicate the transmitter and receiver, respectively.  Specifically, the array response vectors $\boldsymbol{\alpha}_x(\theta_{r,1},\phi_{r,1})$, $\boldsymbol{\alpha}_y(\theta_{r,1},\phi_{r,1})$, $\boldsymbol{\alpha}_y(\theta_{t,1},\phi_{t,1})$,  and $\boldsymbol{\alpha}_z(\phi_{t,1})$ can be written as~\cite{Tsai2018}
 \begin{align}
    & \boldsymbol{\alpha}_x(\theta_{r,1},\phi_{r,1}) = \Big[e^{-j \frac{2\pi  d_{\text{R},x}}{\lambda} \big(\frac{M_{\text{R},x} -1}{2}\big) \cos(\theta_{r,1}) \sin(\phi_{r,1})}, \nonumber\\
    & \cdots, e^{j \frac{2\pi d_{\text{R},x}}{\lambda} \big(\frac{M_{\text{R},x} -1}{2}\big) \cos(\theta_{r,1})\sin(\phi_{r,1})} \Big]^{\mathsf{T}},    \\
    &\boldsymbol{\alpha}_y(\theta_{r,1},\phi_{r,1}) = \Big[e^{-j \frac{2\pi  d_{{\text{R},y}}}{\lambda} \big(\frac{M_{\text{R},y} -1}{2}\big) \sin(\theta_{r,1}) \sin(\phi_{r,1})}, \nonumber\\
    & \cdots, e^{j \frac{2\pi d_{{\text{R},y}}}{\lambda} \big(\frac{M_{{\text{R},y}} -1}{2}\big) \sin(\theta_{r,1})\sin(\phi_{r,1})} \Big]^{\mathsf{T}},\\    &\boldsymbol{\alpha}_y(\theta_{t,1},\phi_{t,1}) = \Big[e^{-j \frac{2\pi  d_{{\text{B},y}}}{\lambda} \big(\frac{M_{{\text{B},y}} -1}{2}\big) \sin(\theta_{t,1}) \sin(\phi_{t,1})}, \nonumber\\
    & \cdots, e^{j \frac{2\pi d_{{\text{B},y}}}{\lambda} \big(\frac{M_{{\text{B},y}} -1}{2}\big) \sin(\theta_{t,1})\sin(\phi_{t,1})} \Big]^{\mathsf{T}},\\
       &\boldsymbol{\alpha}_z(\phi_{t,1}) = \Big[e^{-j \frac{2\pi d_{\text{B},z}}{\lambda} \big(\frac{M_{\text{B},z} -1}{2}\big) \cos(\phi_{t,1}) }, \nonumber\\
    & \cdots, e^{j \frac{2\pi d_{\text{B},z}}{\lambda} \big(\frac{M_{\text{B},z} -1}{2}\big) \cos(\phi_{t,1})} \Big]^{\mathsf{T}},
 \end{align}
 where $M_\text{R} = M_{\text{R},x} M_{\text{R},y}$ with $M_{\text{R},x}$ and $M_{\text{R},y}$ being the number of RIS elements across $x$ axis and $y$ axis, respectively, $d_{\text{R},x}$ and $d_{\text{R},y}$ are the inter-element spacing for the RIS elements across $x$-axis and $y$-axis. Similarly, $M_\text{B} = M_{\text{B},y} M_{\text{B},z}$ with $M_{\text{B},y}$ and $M_{\text{B},z}$ being the number of BS antennas across $y$-axis and $z$-axis, respectively, $d_{\text{B},y}$ and $d_{\text{B},z}$ are the inter-element spacing for the BS antennas across $y$-axis and $z$-axis, respectively. 
 
 Similarly, the remaining channels (marked in Fig.~\ref{System_Model}), e.g., $\bh_2\in\mathbb{C}^{1\times M_\text{B}}$, $\bh_3\in\mathbb{C}^{1\times M_\text{R}}$, $\bh_4 \in\mathbb{C}^{M_\text{U}\times 1}$, and $\bH_5\in\mathbb{C}^{M_\text{U}\times M_\text{B}}$ can be presented in the same manner, listed below:
 \begin{align}
     \bh_2 &=\frac{e^{-j 2\pi  d_2/\lambda}}{\sqrt{\rho_2}} \Big(\boldsymbol{\alpha}_y(\theta_{t,2},\phi_{t,2}) \otimes \boldsymbol{\alpha}_z(\phi_{t,2})\Big)^\mathsf{H},\\
     \bh_3 &=\frac{e^{- j 2\pi   d_3/\lambda}}{\sqrt{\rho_3}} \Big(\boldsymbol{\alpha}_x(\theta_{t,3},\phi_{t,3}) \otimes \boldsymbol{\alpha}_y(\theta_{t,3},\phi_{t,3})\Big)^\mathsf{H},\\
     \bh_4 &=\frac{e^{-j 2\pi  d_4/\lambda}}{\sqrt{\rho_4}} \boldsymbol{\alpha}_x(\theta_{r,4},\phi_{r,4}) \otimes \boldsymbol{\alpha}_y(\theta_{r,4},\phi_{r,4}), \\
     \bH_5 &= \frac{e^{-j 2\pi  d_5/\lambda}}{\sqrt{\rho_5}} \boldsymbol{\alpha}_x(\theta_{r,5},\phi_{r,5}) \otimes \boldsymbol{\alpha}_y(\theta_{r,5},\phi_{r,5})\nonumber\\ &\hspace{0.47cm}\Big(\boldsymbol{\alpha}_y(\theta_{t,5},\phi_{t,5}) \otimes \boldsymbol{\alpha}_z(\phi_{t,5})\Big)^\mathsf{H},
 \end{align}
where $M_\text{U}$ is the number of UE antennas, $d_2(\rho_2)$, $d_3(\rho_3)$, and $d_4(\rho_4)$ are the distance (path loss) between the drone and the BS, the RIS, and the UE, respectively, $d_5(\rho_5)$ is the distance (path loss) between the BS and the UE. $\theta_{t,i}$ and $\phi_{t,i}$ are the azimuth and elevation angles of departure associated with $\bh_i$ for $i = 2,3$, and $\theta_{r,4}$ ($\theta_{r,5}$) and $\phi_{r,4}$ ($\phi_{r,5}$) are the azimuth and elevation angles of arrival associated with $\bh_4$~($\bH_5$). $\theta_{t,5}$ and $\phi_{t,5}$ are the azimuth and elevation angles of departure associated with $\bH_5$. In the drone detection system, $\bH_5$ is the direct channel between the BS and the UE, which acts as an interference link in the studied drone detection problem.  
 
 \subsection{Sounding Procedures} \label{sec_sounding_procedure}
During the sounding stage, the BS sends beamformed pilot signals illuminating the RIS and a certain part of the sky simultaneously, meanwhile the RIS steers the beams to cover a certain range of the sky, where the drone is potentially located. By following this, we can collaboratively cover a larger space of the sky compared to its RIS-free counterpart. The RIS training beam design and steering is coordinated by the BS for all time slots within the sounding period. The direct signal from the BS and the reflected signal from the RIS will arrive at the drone, and then scattered/reflected to the UE, i.e., the signal from the BS to the UE propagates via one single-bounce path (BS-Drone-UE) and one double-bounce path (BS-RIS-Drone-UE). 

\begin{figure}[t]
	\centering
\includegraphics[width=0.8\linewidth]{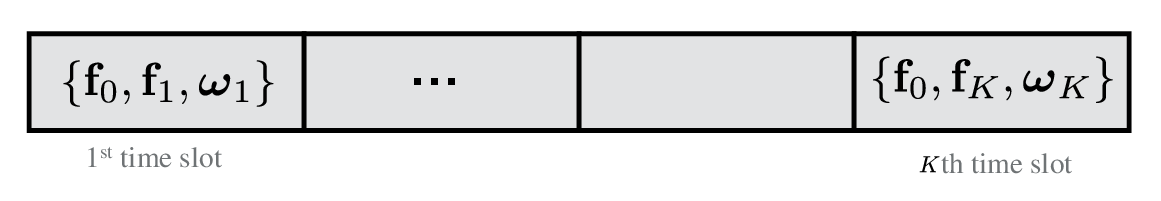}
	\caption{The sounding frame for drone detection, where the beam $\bf_0 \in \mathbb{C}^{M_\text{B} \times 1}$ is directed towards the RIS, and the beam pair $\{\bf_k \in \mathbb{C}^{M_\text{B} \times 1}, \boldsymbol{\omega}_k \in \mathbb{C}^{M_\text{R} \times 1}\}$ are directed towards the sky.}	\label{Sounding_Frame}
 		\vspace{-0.5cm}
\end{figure}

Recall that we assume that the exact position/location of RIS is known to the other network nodes, i.e., BS and UE. In this sense, the BS can take full advantage of such prior information and use it for the purpose of beam design during the sounding process. We further assume that the BS is capable of sending multiple (orthogonal or semi-orthogonal) beams (i.e., spatially multiplexed data streams) simultaneously. For instance, at the $k$th time slot shown in Fig.~\ref{Sounding_Frame}, the BS sends the symbol with beamforming vector $\bf_0$ ($\|\bf_0\|_2 = 1$) towards the RIS and the other symbol with beamforming vector $\bf_k$ ($\|\bf_k\|_2 = 1$) towards the sky, where $\bf_0^\mathsf{H} \bf_k \approx 0 $, for $k = 1,\cdots, K$. Meanwhile, the RIS considers the phase profile $\boldsymbol{\omega}_k$ to beamform/steer the signal to the sky as well. Each element of $\boldsymbol{\omega}_k$ satisfies unit-modulus constraint, i.e., $|[\boldsymbol{\omega}_k]_i| = 1$, for $\forall i$. We further assume that the drone is a point scatterer with reflection coefficient $\zeta$, which is interchangeable with RCS.\footnote{The RCS varies from one drone to another, depending on the size, shape, and material of the drone. Here, without loss of generality, different realistic reflection coefficients are considered for the numerical study in Section~\ref{sec_Numerical_study}.} Other kinds of assumption on the RCS or its distribution can also be found in the literature~\cite{Farlik2016}.

Note that during the sounding procedure, the beam design at the BS and RIS are essential for the drone detection systems. By following MRT principle based on the known positions, we set $\bf_0 = \frac{1}{\sqrt{M_\text{B}}} \big(\boldsymbol{\alpha}_y(\theta_{t,1},\phi_{t,1}) \otimes \boldsymbol{\alpha}_z(\phi_{t,1})\big)$ and define $\bg_0 = \frac{1}{\sqrt{M_\text{B}}} \big(\boldsymbol{\alpha}_y(\theta_{t,5},\phi_{t,5}) \otimes \boldsymbol{\alpha}_z(\phi_{t,5})\big)$, we can make the remaining beams (i.e., $\bf_k$, for $k =1,\cdots, K$) fall into the null space of the matrix $\bF^\mathsf{H}$, where $\bF = [\bf_0 ,\; \bg_0]$. Namely, $\bf_k^\mathsf{H} \bf_0  = \bf_k^\mathsf{H} \bg_0 =0$, $\forall k$.\footnote{We need to assume that $K \leq M_\text{B} -2$. Otherwise, the beams at the BS need to be reused along with a new series of RIS beams.}

\subsection{Signal Model}
The received signal at the UE during the time slot $k$ can be expressed as 
\begin{equation}\label{by_k}
    \by_k =  \zeta\bh_4 (\bh_3 \mathrm{diag}(\boldsymbol{\omega}_k) \bH_1 \bF_k \bs_k + \bh_2 \bF_k \bs_k) + \bH_5 \bF_k \bs_k +\bn_k,
\end{equation}
where $\bF_k = [\bf_0,\, \bf_k]$, $\bs_k = [s_0,\, s_k]^\mathsf{T}$, and the transmitted symbols during each time slot satisfying the following sum power constraint: $\mathbb{E}\{|s_0|^2\} + \mathbb{E}\{|s_k|^2\} = P$ for $\forall k$, and the additive noise vector at the UE follows complex Gaussian distribution, i.e., $\bn_k \sim \mathcal{CN}(\mathbf{0},\sigma^2 \bI_{M_\text{U}})$.
The signal vector $\by_k$ in~\eqref{by_k} can be further expressed as 
\begin{align}\label{by_k1}
    \by_k =&  \zeta\bh_4 (\bh_3\mathrm{diag}(\boldsymbol{\alpha}_x(\theta_{r,1},\phi_{r,1}) \otimes \boldsymbol{\alpha}_y(\theta_{r,1},\phi_{r,1})) \boldsymbol{\omega}_k \nonumber\\ &\times \frac{e^{-j 2\pi  d_1/\lambda}}{\sqrt{\rho_1}} \big(\boldsymbol{\alpha}_y(\theta_{t,1},\phi_{t,1}) \otimes \boldsymbol{\alpha}_z(\phi_{t,1})\big)^\mathsf{H} \bx_k \nonumber\\
    &+ \bh_2 \bx_k) + \bH_5 \bx_k + \bn_k,
\end{align}
where $\bx_k = \bF_k \bs_k$. 
By stacking $\{\by_k\}$ column by column, the received signal can be expressed as 
\begin{equation}\label{Rx_signal_UE}
    \bY = \tilde{\bH} \tilde{\boldsymbol{\Omega}} + \hat{\bH} \bX + \bH_5\bX +  \bN,
\end{equation}
where $\tilde{\boldsymbol{\Omega}} = [ \eta_1 \boldsymbol{\omega}_1, \cdots, \eta_K \boldsymbol{\omega}_K ]$ with $\eta_k = \big(\boldsymbol{\alpha}_y(\theta_{t,1},\phi_{t,1}) \otimes \boldsymbol{\alpha}_z(\phi_{t,1})\big)^\mathsf{H} \bx_k$, $\bX = [\bx_1, \cdots, \bx_K]$, and $\bN = [\bn_1, \cdots, \bn_K]$. Note that $\eta_k$, for $k = 1, \cdots, K$, is known \textit{a priori} since both $\big(\boldsymbol{\alpha}_y(\theta_{t,1},\phi_{t,1}) \otimes \boldsymbol{\alpha}_z(\phi_{t,1})\big)^\mathsf{H}$ and $\bx_k$ are pre-known. The two cascaded channels $ \tilde{\bH}$ and $ \hat{\bH}$ are function of $\zeta$, $\bh_4$, $\bh_3$, $\mathrm{diag}( \boldsymbol{\alpha}_x(\theta_{r,1},\phi_{r,1}) \otimes \boldsymbol{\alpha}_y(\theta_{r,1},\phi_{r,1})  )$, and $\bh_2$, as
\begin{align}\label{H_tilde}
   \tilde{\bH} &=  \zeta  \frac{e^{-j 2\pi  d_1/\lambda}}{\sqrt{\rho_1}} \bh_4 \bh_3\mathrm{diag}(\boldsymbol{\alpha}_x(\theta_{r,1},\phi_{r,1}) \otimes \boldsymbol{\alpha}_y(\theta_{r,1},\phi_{r,1})), \\
   \hat{\bH} &=\zeta\bh_4\bh_2  \nonumber\\
   &= \hat{\epsilon} \boldsymbol{\alpha}_x(\theta_{r,4},\phi_{r,4}) \boldsymbol{\alpha}_y(\theta_{t,2},\phi_{t,2})^\mathsf{H} \otimes \boldsymbol{\alpha}_y(\theta_{r,4},\phi_{r,4}) \boldsymbol{\alpha}_z(\phi_{t,2})^\mathsf{H}, 
\end{align}
where $\hat{\epsilon} =\zeta \frac{e^{-2\pi d_4/\lambda}}{\sqrt{\rho_4}} \frac{e^{-2\pi d_2/\lambda}}{\sqrt{\rho_2}}$.

Since we can assume that the angular parameters in $\bH_1$ are known \textit{a priori}, $\mathrm{diag}(\boldsymbol{\alpha}_x(\theta_{r,1},\phi_{r,1}) \otimes \boldsymbol{\alpha}_y(\theta_{r,1},\phi_{r,1}))$ is known in $\tilde{\bH}$. The unknown parameters in $\tilde{\bH}$ are $\bh_3$, $\bh_4$, $\zeta$, and $\frac{e^{- j 2\pi d_1/\lambda}}{\sqrt{\rho_1}}$. In~$\hat{\bH}$, all the three terms, i.e., $\zeta$, $\bh_4$, and $\bh_2$, are unknown.  
Recall that the goal for the study is to perform the passive detection task on the absence/presence of the drone based on the received signal $\bY$ in~\eqref{Rx_signal_UE}. 


\section{Drone Detection via GLRT}
\subsection{Derivation of Probability of Detection}
For the drone detection, we resort to the GLRT, which is detailed below~\cite{kay1998statistical}
\begin{equation}
L(\bY) = \frac{p(\bY;\mathcal{H}_1)}{p(\bY;\mathcal{H}_0)}> \gamma, 
\end{equation}
where the threshold value $\gamma$ is computed from the pre-set probability of false alarm:
\begin{equation}
P_\text{FA} = \int_{\{\bY: L(\bY) >\gamma\}} p(\bY;\mathcal{H}_0) d\bY = \alpha, 
\end{equation}
and the two hypotheses (i.e., absence and presence of the drone) are 
\begin{align}
&\mathcal{H}_0 :     \bY = \bH_5\bX+ \bN, \\
&\mathcal{H}_1 :     \bY =  \tilde{\bH} \tilde{\boldsymbol{\Omega}} + \hat{\bH} \bX + \bH_5\bX+ \bN,
\end{align}
where $p(\bY;\mathcal{H}_0)$ and $p(\bY;\mathcal{H}_1)$ are the probability density functions (pdfs) under these hypotheses. 
For the ease of analytical simplicity, we perform vectorization of the received signal under the two aforementioned hypotheses. The expressions are 
\begin{align}
\mathcal{H}_0 :     \by &=(\bX ^\mathsf{T}\otimes \bI_{M_\text{U}})  \bh_5 +   \bn, \\
\mathcal{H}_1 :     \by &=  (\tilde{\boldsymbol{\Omega}} ^\mathsf{T}\otimes \bI_{M_\text{U}}) \tilde{\bh}  +  (\bX ^\mathsf{T}\otimes \bI_{M_\text{U}})  \hat{\bh} \nonumber\\
&+ (\bX ^\mathsf{T}\otimes \bI_{M_\text{U}})  \bh_5+ \bn, 
\end{align}
where $\by = \mathrm{vec}(\bY)$, $\bn = \mathrm{vec}(\bN)$, $\tilde{\bh} = \mathrm{vec}(\tilde{\bH})$, $\hat{\bh} = \mathrm{vec}(\hat{\bH})$, and $\hat{\bh}_5 = \mathrm{vec}(\hat{\bH}_5)$. We introduce $\bz = (\bX ^\mathsf{T}\otimes \bI_{M_\text{U}})  \bh_5 +   \bn$, which is the interference term plus noise. It is distributed as $\mathcal{CN} ((\bX ^\mathsf{T}\otimes \bI_{M_\text{U}})  \bh_5, \sigma^2 \bI_{M_\text{U}} + (\bX ^\mathsf{T}\otimes \bI_{M_\text{U}})  \bh_5 \bh_5^\mathsf{H}(\bX ^\mathsf{T}\otimes \bI_{M_\text{U}})^\mathsf{H} )$, which can be regarded as colored noise.\footnote{Since the channel $\bH_5$ is stationary regardless of the absence/presence of the drone, so we can get the first-order and second-order statistics at the initial phase and use them in the sequel drone detection phase.} The two hypotheses can be further reformulated by following the well-known pre-whitening process~\cite{kay1998statistical}, as 
\begin{align}
 &\mathcal{H}_0 :  \tilde{\by} = \bR  \bz- \bR\boldsymbol{\mu}, \\
  &\mathcal{H}_1 :  \tilde{\by} = \bR \boldsymbol{\Psi} \bh + \bR \bz - \bR \boldsymbol{\mu},
\end{align}
where $\boldsymbol{\Psi} = \Big[(\tilde{\boldsymbol{\Omega}} ^\mathsf{T}\otimes \bI_{M_\text{U}}), \; (\bX ^\mathsf{T}\otimes \bI_{M_\text{U}}) \Big]$, $\bh = [ \tilde{\bh}^\mathsf{T}, \; \hat{\bh}^\mathsf{T} ]^\mathsf{T}$, $\boldsymbol{\mu}  =  (\bX ^\mathsf{T}\otimes \bI_{M_\text{U}})  \bh_5$, and the Cholesky decomposition/factorization of $\bC^{-1}$ is $\bC^{-1} = \bR^\mathsf{H}\bR$ with $\bC = \sigma^2 \bI_{M_\text{U}} + (\bX ^\mathsf{T}\otimes \bI_{M_\text{U}})  \bh_5 \bh_5^\mathsf{H}(\bX ^\mathsf{T}\otimes \bI_{M_\text{U}})^\mathsf{H}$. 

The pdf of $\tilde{\by}$ under $\mathcal{H}_0$, $p(\tilde{\by};\mathcal{H}_0)$, is given by 
\begin{equation}
    p(\tilde{\by};\mathcal{H}_0) = \frac{1}{(\pi )^{K M_\text{U}}} \exp(- \tilde{\by}^\mathsf{H} \tilde{\by}).
\end{equation}
The pdf of $\tilde{\by}$ under $\mathcal{H}_1$ depends on the unknown parameter $\bh$ (a tandem combination of two unknown parameters $\tilde{\bh}$ and $\hat{\bh}$), and it is written as 
\begin{equation} 
    p(\tilde{\by}|\bh;\mathcal{H}_1) = \frac{1}{(\pi)^{K M_\text{U}}} \exp\big(- (\tilde{\by} - \bR\boldsymbol{\Psi}\bh)^\mathsf{H} (\tilde{\by} - \bR\boldsymbol{\Psi}\bh )  \big).
\end{equation}
Following the GLRT principle, we decide $\mathcal{H}_1$ if 
\begin{equation}
    L(\by) = \frac{\exp\big(- (\tilde{\by} - \bR\boldsymbol{\Psi}\breve{\bh} )^\mathsf{H} (\tilde{\by} - \bR\boldsymbol{\Psi}\breve{\bh} )  \big)}{\exp(- \tilde{\by}^\mathsf{H} \tilde{\by} )} > \gamma,
\end{equation}
where $\breve{\bh}$ is the maximum likelihood (ML) estimate of $\bh$, expressed as
\begin{equation}
 \breve{\bh}  = (\bR\boldsymbol{\Psi})^\dag \tilde{\by}.
\end{equation}
After taking the natural logarithm, we have 
\begin{align}
    2\ln(L(\by)) &= 2(\|\tilde{\by}\|^2 -\|\by -\bR\boldsymbol{\Psi} (\bR\boldsymbol{\Psi})^\dag \tilde{\by}\|^2  ) \nonumber\\
    & = \frac{\|\bR \boldsymbol{\Psi}\breve{\bh} \|^2}{1/2}  >2\ln(\gamma) = \gamma^\prime.
\end{align}
The probability of false alarm is 
\begin{equation}
    P_\text{FA} = Q_{\chi^2_{2M_\text{U} K}} (\gamma^\prime),
\end{equation}
where $Q_{\chi^2_{2M_\text{U} K}}(\cdot)$ is the right-tail probability for a Chi-squared random variable with degree of freedom being $2M_\text{U} K$, denoted by $\chi^2_{2M_\text{U} K}$~\cite{kay1998statistical}.  The probability of detection is 
\begin{equation}\label{eq:P_D}
    P_\text{D} = Q_{\chi^{\prime2}_{2M_\text{U} K}(\lambda)} (\gamma^\prime),
\end{equation}
where $\chi^{\prime2}_{2M_\text{U} K}(\lambda)$ is the noncentral Chi-squared distribution with noncentrality parameter $\lambda  = 2\|\bR \boldsymbol{\Psi} \bh \|^2$ and $Q_{\chi^{\prime2}_{2M_\text{U} K}(\lambda)} (\cdot)$ is the right-tail probability for a random variable distributed as $\chi^{\prime2}_{2M_\text{U} K}(\lambda)$.

\subsection{RIS Training Beam Design}
According to the expression of $P_\text{D}$ in Eq.~\eqref{eq:P_D}, with fixed degree of freedom and threshold value $\gamma^\prime$, in order to maximize $P_\text{D}$, we need to maximize $\lambda$. The reason lies in that $P_\text{D}$ monotonically increases as the increase of $\lambda$ when the other parameters, i.e., degree of freedom and $\gamma^\prime$, are fixed. The proof is included in the Appendix. The optimization of the RIS training beam matrix design can be formulated as 
\begin{align}\label{Eq_RIS_opt_design}
    \max_{\{\boldsymbol{\omega}_k\}} &\;\;\lambda (\bX, \tilde{\boldsymbol{\Omega}}) = \bh^\mathsf{H} \boldsymbol{\Psi}^\mathsf{H} \bC \boldsymbol{\Psi} \bh  \\
    \text{s.t.}\; &\;\;\mathrm{Tr}(\bX\bX^\mathsf{H}) = KP, \nonumber \\
    &\;\;\mathrm{Tr}(\tilde{\boldsymbol{\Omega}}\tilde{\boldsymbol{\Omega}}^\mathsf{H}) = KP M_\text{B} M_\text{R}/2,\nonumber \\
    &\;\; |[\boldsymbol{\omega}_k]_i| = 1, \forall k, i.\nonumber 
\end{align}
In $\tilde{\boldsymbol{\Omega}}$, $\boldsymbol{\omega}_k$ and $ \bx_k$ are coupled, and $\bx_1, \cdots, \bx_K$ are mutually coupled. In addition, the last constraint $|[\boldsymbol{\omega}_k]_i| = 1$ is non-convex. Therefore, the optimization problem in Eq.~\eqref{Eq_RIS_opt_design} is difficult to be addressed. Finding the optimal solution to the problem~\eqref{Eq_RIS_opt_design} will be left for our future investigation. Instead, we will select three exemplary RIS training beam designs and study their effect on probability of drone detection in Section~\ref{sec_Numerical_study}. 

\section{Numerical Results and Discussion}\label{sec_Numerical_study}
In this section, we evaluate the performance of the RIS-augmented mmWave MIMO drone detection systems. The parameters are set as follows: $M_\text{B} = 10 \times 10 =  100$, $M_\text{R}= 40 \times 40 = 1600$, $M_\text{U} = 4\times 4 = 16$, and the carrier frequency $f_c$ is set to be 28 GHz. The BS is located at $(0,0,28)$, the RIS is located at $(0.1,0.1,27.9)$, the drone is located at $(1,1,29.5)$, and the UE is located at $(2,2, 27)$, all in meters. By following the geometric relationship, we can calculate the angular parameters for all the five individual channels. The channel bandwidth $B$ is set to $10$ MHz, and noise variance $\sigma^2 = -174\; \text{dBm} + 10 \log_{10}(B) = -104$ dBm. The transmit power $P$ at the BS is set from $20$ dBm to $40$ dBm. We only account for the free-space path loss (in dB), following Friis’ Law, as follows:
\begin{equation}
    \rho_i (d_i) = 20 \log10 (d_i) - 87.55 + 20 \log10(f_c), \; \forall i = 1,\ldots,5, 
\end{equation}
where $d_i$ is the distance in meters and $f_c$ is measured in KHz. 

\subsection{Benchmarks}
The most intuitive and straightforward benchmark is to consider the drone detection without the aid of an RIS. In this regard, we only consider the single-bound propagation path in the RIS-free drone detection systems. 

\begin{figure}[t]
	\centering
\includegraphics[width=0.75\linewidth]{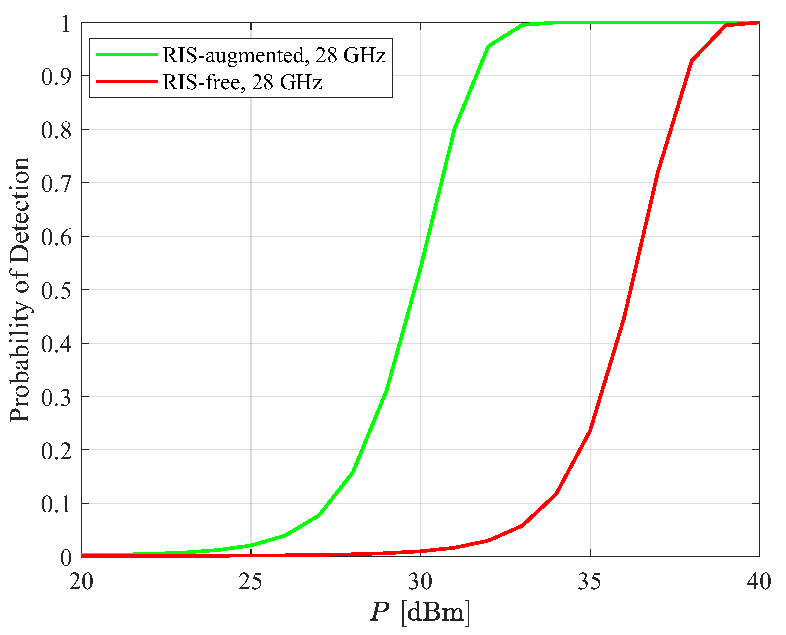}
	\caption{Probability of detection performance with and without using an RIS,  where $K = 90$, $P_\text{FA} = 0.001$, and RIS phase profiles are randomly generated.}
		\label{Random_RISvsRISfree}
 		\vspace{-0.5cm}
\end{figure}

Without the aid of RIS, the two hypotheses are 
\begin{align}
&\mathcal{H}_0 :     \bY = \bH_5\bX+ \bN, \\
&\mathcal{H}_1 :     \bY =   \hat{\bH} \bX + \bH_5\bX+ \bN. 
\end{align}
Therefore, the probability of detection for the RIS-free system is 
\begin{equation}
    \bar{P}_\text{D} = Q_{\chi^{\prime2}_{2M_\text{U} K}(\bar{\lambda})} (\gamma^\prime),
\end{equation}
where $\bar{\lambda}  = 2\|\bR (\bX ^\mathsf{T}\otimes \bI_{M_\text{U}})  \hat{\bh} \|^2$. The drone detection results for the RIS-augmented and RIS-free mmWave MIMO systems are depicted in Fig.~\ref{Random_RISvsRISfree} with $K=90$, $P_\text{FA} = 0.001$, and random RIS phase profile design. The performance gain of RIS-augmented system exceeds $5$ dB for a certain value of probability of detection.

\subsection{Effect of RIS Training Beams}
\begin{figure}[t]
	\centering
\includegraphics[width=0.75\linewidth]{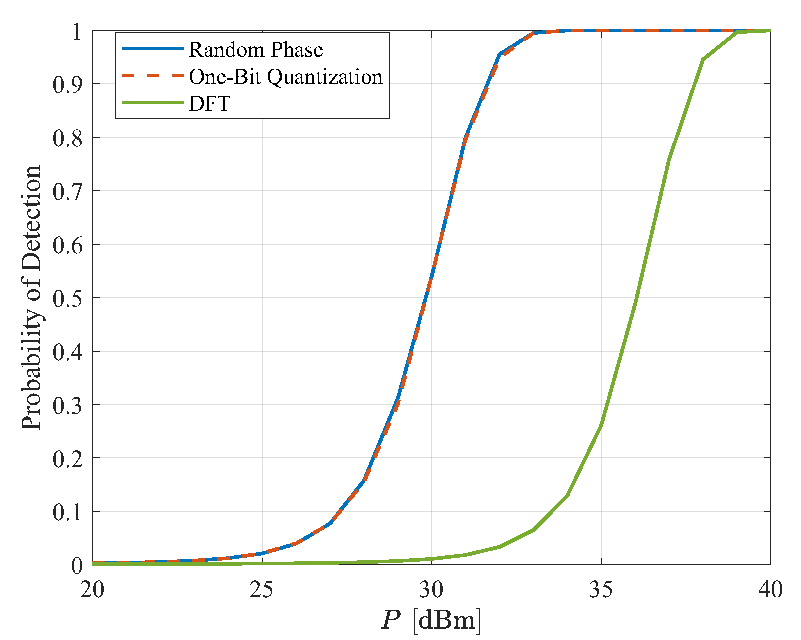}
	\caption{Probability of detection performance with different kinds of RIS training beam designs, where $K = 90$, $P_\text{FA} = 0.001$, and $\zeta = 0.3$.}
		\label{Drone_Detection_training_beam}
 		\vspace{-0.5cm}
\end{figure}

We fix the beam design at the BS, as stated in Section~\ref{sec_sounding_procedure}, while considering different beam design methods for the RIS. The training beams at the RIS are designed as follows: (i) random beams (each element of the beam vectors is constant modulus with random phase), (ii) low phase resolution, e.g., one-bit quantization (i.e., two phase choices, 0 and $\pi$, are used for each RIS phase shifter), (iii) the RIS training beam training matrix is part of a discrete Fourier transform (DFT) matrix. The simulation results are shown in Fig.~\ref{Drone_Detection_training_beam}, from which we observe that the random design has similar performance with one-bit quantization while outperforming the DFT scheme. It is because that the DFT design loses the randomness nature of the RIS beams for the drone detection task.  

\subsection{Effect of Training Overhead}
\begin{figure}[t]
	\centering
\includegraphics[width=0.75\linewidth]{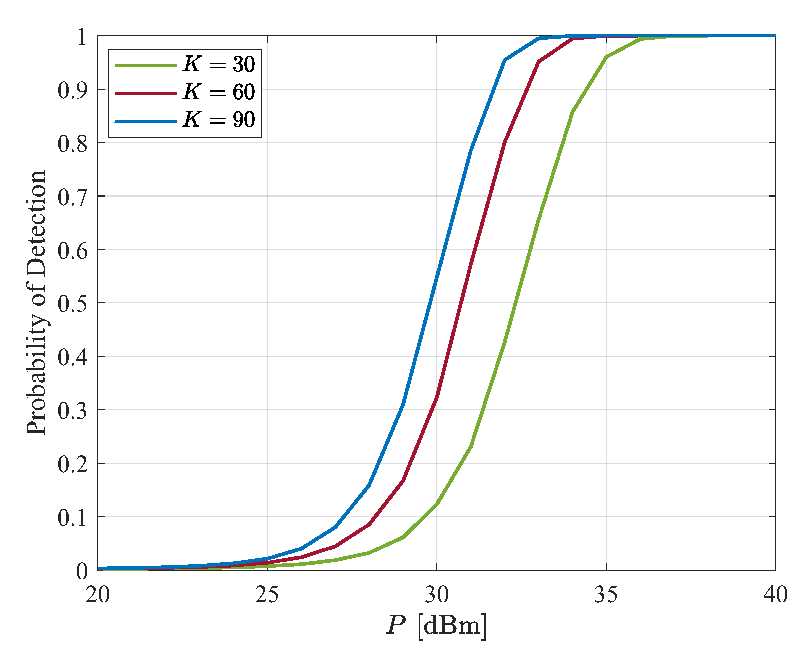}
	\caption{Probability of detection performance with different $K$ values and random RIS training beams, i.e., $K \in \{30,60,90\}$, where $P_\text{FA} = 0.001$, $\zeta = 0.3$.}
		\label{Drone_Detection_Training_Overhead}
 		\vspace{-0.5cm}
\end{figure}

We evaluate the effect of training overhead on the detection performance. Different training overheads, i.e., $K$ values, are considered. The simulation results are presented in Fig.~\ref{Drone_Detection_Training_Overhead}. As expected, the higher the training overhead, the better the performance. However, the increase is not so significant when increasing $K$ by $30$. In other words, we can perform efficient drone detection with a small amount of training overhead in practice.  

\subsection{Effect of RCS}
\begin{figure}[t]
	\centering
\includegraphics[width=0.75\linewidth]{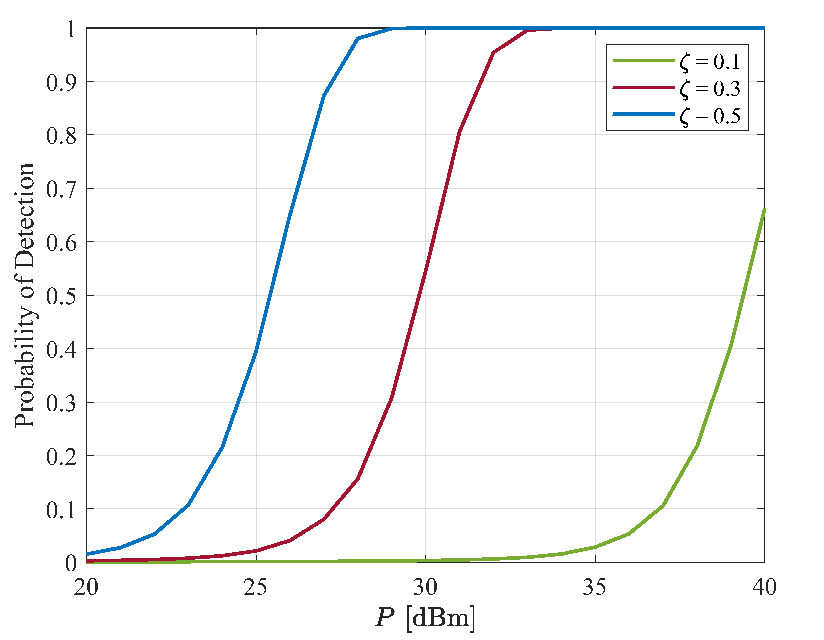}
	\caption{Probability of detection performance with different $\zeta$ values, i.e., $\zeta \in \{0.1, 0.3 , 0.5\}$, and random RIS training beams, where $K = 90$ and $P_\text{FA} = 0.001$.}
		\label{Drone_Detection_effect_RCS}
 		\vspace{-0.5cm}
\end{figure}

The RCS values affect the drone detection performance. Thus, we evaluate the drone detection performance with different RCS values, e.g., $\zeta \in \{ 0.1, 0.3, 0.5\}$~\cite{Semkin2020}. The simulation results are shown in Fig.~\ref{Drone_Detection_effect_RCS}. The RCS values significantly affect the received power strength, which in turns affect the drone detection performance. The performance gap between $\zeta =0.1$ and $\zeta = 0.3$ is around $10$ dB at the level of probability of detection equal to $0.7$, while the gap is reduced to around half when extending $\zeta$ from $\zeta = 0.3$ to $\zeta = 0.5$. 

\section{Conclusion}
In this paper, we have presented an RIS-augmented mmWave MIMO drone detection system in a passive device-free fashion, which can rely on the existing cellular infrastructure. We have verified that the proposed system can outperform its RIS-free counterpart over mmWave frequency bands. In addition, the effect of the RIS training beam design, the training overhead, and RCS, on the detection performance has been investigated to offer practical insights for the proposed device-free drone detection system. 

We further claim that the RIS training beams used during the detection process can be further optimized if some rough prior information about the drone is available, and we left this research topic as our future work. In addition, the RCS value will be in general dependent on the angle of incidence, and this effect will be examined in a future work. 

\section*{Acknowledgement}
Aymen Fakhreddine's contribution is partly funded by the Austrian Science Fund (FWF\,--\,Der Wissenschaftsfonds) under grant ESPRIT-54 (Grant DOI: 10.55776/ESP54). The work of Prof. Alexandropoulos has been supported by the SNS JU project
6G-DISAC under the EU's Horizon Europe research and
innovation programme under Grant Agreement No 101139130.

\section*{Appendix}
\subsection*{Monotonicity of $P_\text{D}$ with respect to $\lambda$}
To prove that the right-tail probability of the noncentral Chi-squared distribution increases monotonically as a function of $\lambda$, we show instead that its cumulative distribution function (cdf) decreases monotonically as a function of $\lambda$. In particular, the cdf can be expressed as follows:
\begin{equation}
    F(x;k,\lambda) = e^{-\lambda/2} \sum_{l=0}^{\infty} \frac{(\lambda/2)^l}{l !} F(x, k+2l),
\end{equation}
where $F(x, k + 2l)$ is the cdf of the central Chi-squared distribution with $k$ degrees of freedom. We compute the first-order derivative $\partial F(x;k,\lambda)/ \partial \lambda$ as: 
\begin{equation}
    \frac{\partial F(x;k,\lambda)}{ \partial \lambda}\!=\!\frac{1}{2} e^{\lambda/2} \sum_{l = 0}^{\infty} \frac{ (\lambda/2)^l}{l !}[F(x; k+2l+2) - F(x; k+2l)].
\end{equation}
Next, we prove that $F(x; k + 2l + 2) - F(x; k+ 2l) <0$, $\forall l$. Namely, we calculate the following difference:
\begin{equation}\label{CDF_difference}
    F(x; k + 2) - F(x;k) = \frac{\gamma(k/2 +1, x/2)}{\Gamma(k/2 +1)} - \frac{\gamma(k/2, x/2)}{\Gamma(k/2 )},
\end{equation}
where $\gamma(\cdot)$ is the lower incomplete gamma function and $\Gamma(\cdot)$ is the ordinary gamma function. By using the power series expansion of $\gamma(s,x) = x^s \Gamma(s) e^{-x} \sum\limits_{l=0}^{\infty} \frac{x^l}{\Gamma(s+l+1)}$, Eq. \eqref{CDF_difference} can be further expressed as follows:
\begin{align}
  &F(x; k + 2) - F(x;k) =  \frac{\gamma(k/2 +1, x/2)}{\Gamma(k/2 +1)} - \frac{\gamma(k/2, x/2)}{\Gamma(k/2 )} \nonumber\\
  & = \left(\frac{x}{2}\right)^{k/2 +1} e^{-\frac{x}{2}} \sum_{l=0}^{\infty} \frac{\left(\frac{x}{2}\right)^l}{ \Gamma(k/2+l+2)} \nonumber\\ &\hspace{0.4cm}-\left(\frac{x}{2}\right)^{k/2} e^{-\frac{x}{2}} \sum_{l=0}^{\infty} \frac{\left(\frac{x}{2}\right)^l}{ \Gamma(k/2+l+1)}  = - \frac{ \left(\frac{x}{2}\right)^{k/2} e^{-\frac{x}{2}}}{ \Gamma(k/2 +1)} < 0.
\end{align}
Thus, $\partial F(x;k,\lambda)/ \partial \lambda < 0$, $\forall \lambda$. This completes the proof of the claim that $F(x;k,\lambda)$ decreases monotonically with increasing $\lambda$ values.

\bibliographystyle{IEEEtran}
\bibliography{IEEEabrv,Ref}

\end{document}